\documentclass[preprint2]{aastex}

\slugcomment{The Astrophysical Journal, Submitted 2002 February 20}

\shortauthors{Wanajo et al.}
\shorttitle{$r$-PROCESS AND U-Th CHRONOLOGY}

\begin{document}

\title{The $r$-process in the neutrino winds of core-collapse supernovae and
 U-Th cosmochronology}

\author{Shinya Wanajo$^1$, Naoki Itoh$^1$, Yuhri Ishimaru$^2$, Satoshi
Nozawa$^3$, and Timothy C. Beers$^4$}
\affil{$^1$Department of Physics, Sophia University,
       7-1 Kioi-cho, Chiyoda-ku, Tokyo, 102-8554, Japan;\\
       wanajo@sophia.ac.jp, n\_itoh@sophia.ac.jp\\
       $^2$Institut d'Astrophysique de Paris,
       98 bis, Boulevard Arago, F-75014, Paris, France;
       ishimaru@iap.fr\\
       $^3$Josai Junior College for Women, 1-1 Keyakidai, Sakado-shi,
       Saitama, 350-0290, Japan; snozawa@venus.josai.ac.jp\\
       $^4$Department of Physics/Astronomy, Michigan State University,
       E. Lansing, MI 48824, USA; beers@pa.msu.edu}

\bigskip
\affil{The Astrophysical Journal, submitted 2002 February 20;
                                  accepted 2002 June 9}

\begin{abstract}
The discovery of the second highly $r$-process-enhanced, extremely metal-poor
star, CS~31082-001 ([Fe/H] $= -2.9$) has provided a powerful new tool for age
determination, by virtue of the detection and measurement of the radioactive
species uranium and thorium.  Because the half-life of $^{238}$U is one-third
that of $^{232}$Th, the U-Th pair can, in principle, provide a far more precise
cosmochronometer than the Th-Eu pair that has been used in previous
investigations.  In the application of this chronometer, the age of (the
progenitor of) CS~31082-001 can be regarded as the minimum age of the Galaxy,
and hence of the universe.

One of the serious limitations of this approach, however, is that predictions
of the production ratio of U and Th have not been made in the context of a
realistic astrophysical model of the $r$-process. We have endeavored to produce
such a model, based on the ``neutrino winds'' that are expected to arise from
the nascent neutron star of a core-collapse supernova.  In this model, the
proto-neutron star mass and the (asymptotic) neutrino sphere radius are assumed
to be $2.0 M_\odot$ and 10~km, respectively.  Recent hydrodynamic studies
indicate that there may exist difficulties in obtaining such a compact (massive
and/or small in radius) remnant.  Nevertheless, we utilize this set of
parameter choices since previous work suggests that the third $r$-process peak
(and thus U and Th) is hardly reached when one adopts a less compact
proto-neutron star in the framework of the neutrino-wind scenario.  The
temperature and density histories of the material involved in the
neutron-capture processes are obtained with the assumption of a steady flow of
the neutrino-powered winds, with general relativistic effects taken into
account. The electron fraction is taken to be a free parameter, constant with
time. The $r$-process nucleosynthesis in these trajectories is calculated with
a nuclear reaction network code including actinides up to $Z = 100$.

The mass-integrated $r$-process yields, obtained by assuming a simple time
evolution of the neutrino luminosity, are compared to the available
spectroscopic elemental abundance data of CS~31082-001. As a result, the
``age'' of this star is determined to be $14.1 \pm 2.5$~Gyr, in excellent
agreement with lower limits on the age of the universe estimated by other
dating techniques, as well as with other stellar radioactive age estimates.
Future measurements of Pt and Pb in this star, as well as expansion of searches
for additional $r$-process-enhanced, metal-poor stars (especially those in
which both U and Th are measurable), are of special importance to constrain the
current astrophysical models for the $r$-process.
\end{abstract}


\section{INTRODUCTION}

Uranium and thorium are regarded as potentially useful cosmochronometers
because their long radioactive decay half-lives ($^{232}$Th: 14.05 Gyr,
$^{238}$U: 4.468 Gyr) are significant fractions of the expected age of the
universe.  All of the actinides, including uranium and thorium, are thought to
be pure $r$-process elements, since the alternative neutron-capture process,
the $s$-process, terminates at $^{209}$Bi. The excellent agreement of the
relative abundances of neutron-capture elements in the extremely metal-poor
(${\rm [Fe/H]} = - 3.1$), $r$-process-element-enhanced star CS~22892-052 with
the solar $r$-process pattern initially suggested that thorium might serve as a
precise cosmochronometer \citep{Sned96}. The time that has passed since the
production of the thorium that is now observed in the outer atmosphere of such
an old halo star can be regarded as a hard lower limit on the age of the
universe.\footnote{For convenience of nomenclature, in the context of
radioactive cosmochronology it is common practice to refer to the ``age'' of a
given star as the time difference between the production event (or events) of
the radioactive (and stable $r$-process) species that have been incorporated
into the outer atmosphere of a low-mass ($< 0.8 M_\odot$) star that has not yet
evolved past the giant branch stage of evolution, and hence is amenable to
spectroscopic study.}  Sneden et al. (1996) estimated the age of this star to
be $15.2 \pm 3.7$~Gyr, comparing the measured Th/Eu ratio with an initial
production ratio constrained by theory. This age limit, albeit with a rather
large error bar, is in good agreement with the age limits derived by completely
different techniques, e.g., for globular clusters ($12.9 \pm 2.9$~Gyr; Carretta
et al. 2000), for Type Ia supernovae ($14.9 \pm 1.4$~Gyr; Perlmutter et al.
1999), and from the shape of acoustic peaks in the cosmic microwave background
spectrum ($14.0 \pm 0.5$~Gyr; Knox, Christensen, \& Skordis 2001).

One advantage of the thorium chronology is that, once the initial and current
values of Th/$r$ in the star are specified, the age of the star depends only on
the half-life of $^{232}$Th determined in the laboratory.  That is, one is not
forced to invoke complex (and no doubt incomplete) models of Galactic chemical
evolution, although insight from making simple attempts can still be obtained
(e.g., Cowan et al. 1997). Here, $r$ denotes one of the stable $r$-process
elements, often taken to be Eu. In the most simple model, the initial value of
Th/$r$ may be regarded as that arising from a single supernova that exploded in
the near vicinity (and perhaps may have even triggered the formation of) the
presently observed star, e.g., CS~22892-052, early in the history of the
Galaxy.  In fact, the large dispersion of Eu/Fe observed in halo stars (more
than 2 orders of magnitude)  has been naturally explained by chemical evolution
models that make such assumptions
\citep{Ishi99,Tsuj00}.

Thus far, the initial production of Th/$r$ has been determined by fitting
theoretical nucleosynthesis results to the solar $r$-process pattern, with the
assumption that the $r$-pattern was {\it universal} in all astrophysical
environments \citep{Cowa99,Gori99}. However, a recent abundance analysis of
CS~22892-052 with Keck~I (HIRES) has shown that the lighter neutron-capture
elements ($Z \le 48$) are distinctly underabundant with respect to the solar
pattern, which is well fit by species with $Z \ge 56$ \citep{Sned00}. This
implies that there likely exists at least two $r$-process sites that must be
invoked to account for the entire set of elements associated with the
$r$-process \citep{Ishi00,Qian00}.  Furthermore, it should be noted that
although Johnson \& Bolte (2001) have derived an estimate of the minimum age of
the universe, $11.0 \pm 4.2$ Gyr, based on the abundances of thorium and other
stable neutron-capture elements in five metal-poor stars, the estimated ages of
these stars are very widely spread, from $\approx 6$ to 18~Gyr (with
overlapping error bars).  In addition, recent spectroscopic studies of several
extremely metal-poor halo stars with SUBARU (HDS) has revealed that the spread
of estimated chronometric ages from the Th-Eu pair appears to exceed the
abundance uncertainties (Honda et al.  2002, in preparation).  Taken at face
value, these results imply that the universality of the $r$-pattern may not
hold over the actinides either.

The second discovered $r$-process-enhanced, extremely metal-poor star,
CS~31082-001 (${\rm [Fe/H]} = -2.9$) has provided a new, potentially quite
powerful cosmochronometer, uranium \citep{Cayr01,Hill02}. In principle, uranium
might be expected to be a more precise chronometer than thorium owing to its
shorter half life.  Cayrel et al. (2001) concluded that the age of this star is
$12.5 \pm 3$~Gyr, based on comparisons of the observed ratios of U/Th, U/Os,
and U/Ir with their estimated production ratios \citep{Cowa99,Gori99}. The
discovery of uranium in CS~31082-001 has also prompted additional laboratory
work to refine the oscillator strengths, for both uranium and thorium (Nilsson
et al. 2002a, b), that are required to turn spectroscopic observations into
measured abundances.  However, analysis of this star has also provided
conclusive evidence that the $r$-process pattern is {\it not universal} over
the actinides; the ratio Th/Eu in this star is {\it higher} than that of the
solar $r$-process ratio.  CS~31082-001 would be younger than the solar system
if the initial Th/Eu were taken to be universal, in conflict with the ages
derived from ratios involving U and other stable $r$-process elements close to
the third $r$-process peak.  Therefore, any age estimates that demand
assumption of the universality of the $r$-process pattern may in fact be
unreliable.

In addition to the above non-universality problem, the initial $r$-process
pattern has thus far been determined theoretically by the superposition of
nucleosynthesis results, where one is forced to assume constant temperatures,
neutron number densities, and exposure times (or the number of captured
neutrons) -- the most recent exploration of this approach is presented in
Schatz et al. (2002).  These approximations have been necessary because of the
lack of a reliable astrophysical model for the $r$-process site (at least that
associated with the heavy $r$-process elements). At the moment, the
neutrino-wind scenario is considered to be the most promising, ever since the
work by Woosley et al. (1994), although alternative models involving
neutron-star mergers have also shown some promise \citep{Frei99b}. However,
even the neutrino-wind model encounters serious problems, e.g., the high
entropy ($\sim 400 N_A k$) that led to the robust $r$-processing in the Woosley
et al.  scenario has not been duplicated by other work
\citep{Taka94,Qian96}.

The purpose of this paper is, therefore, to construct a realistic $r$-process
model based on the neutrino-wind scenario, in order to derive initial
production ratios that might be useful for estimation of the minimum age of the
universe, in particular by use of the U-Th chronometer pair.  Recently, Wanajo
et al.  (2001) have demonstrated that the solar $r$-process pattern for $A
\approx 120-200$ is well reproduced in such a model from a compact
proto-neutron star, owing to the inclusion of general relativistic effects.
Although recent hydrodynamic studies have encountered difficulties attaining
the required compactness, we use their model along with some improvements to be
discussed in \S~2.  Although some severe problems yet remain, it is clearly
important to determine the initial $r$-process pattern based on
site-specific, rather than only site-independent, models. In \S~3 the
$r$-process nucleosynthesis calculation is described, and the results are
compared to the solar $r$-process pattern.  The initial electron fraction,
$Y_e$, is varied from 0.39 to 0.49, but taken constant with time during the
neutron-capture process.  In \S~4, the age of CS~31082-001 is derived by
comparison of the nucleosynthesis results obtained for each $Y_e$ with the
abundance pattern of heavy elements in this star.  The implications of this
study are discussed in \S 5.

\section{NEUTRINO WINDS}

Neutrino winds, in which the free nucleons accelerated by the intense neutrino
flux near the neutrino sphere of a core-collapse supernovae assemble to heavier
nuclei, has been believed to be the most promising astrophysical site of the
$r$-process since the work by Woosley et al. (1994).  The rather high entropy
($\sim 400$ $N_A k$) of the ejected matter in their hydrodynamic simulations
led to production of the $r$-process elements that is in remarkable agreement
with the solar $r$-process pattern.  However, subsequent analysis of both
hydrodynamic and analytic approaches suggests that such high entropy may not be
achieved in neutrino winds ($\sim 100$ $N_A k$; Takahashi et al. 1994; Qian \&
Woosley 1996). In addition, an unacceptable overproduction of the elements with
$A \approx 90$ appeared in the $r$-process yields of Woosley et al.  (1994).

Recently, Otsuki et al. (2000) have studied the physical conditions required
for the $r$-process in detail, using a semi-analytic model of a spherical,
steady neutrino wind, taking general relativistic effects into account (see
also Qian \& Woosley 1996; Cardall \& Fuller 1997; Thompson, Burrows, \& Meyer
2001). They suggested that robust physical conditions for the $r$-process were
obtained only if the proto-neutron star was as massive as $\sim 2.0 M_\odot$.
Furthermore, Wanajo et al. (2001) showed from nucleosynthesis calculations that
the $r$-process yields for the compact proto-neutron star models (with a mass
of $1.9 - 2.0 M_\odot$ and radius of 10 km) were in good agreement with the
solar $r$-process pattern. This agreement was achieved due to the short
dynamical timescale ($\lesssim 10$ msec) and moderately high entropy ($\gtrsim
140$ $N_A k$) obtained at the early phase of the neutrino wind, which prevents
most of the free nucleons to assemble into heavy nuclei. There are, however,
some severe problems in their models. For example, it would appear difficult to
obtain such a compact proto-neutron star with hydrodynamic models of
core-collapse supernovae using the equations of state of high-density matter
that are currently available.  This model also suffered from significant
overproduction of elements with $A \approx 90$.  Hence, although problems
remain, we have chosen to begin with these models, and attempt to refine them
in order to see if progress can be made toward the goal of obtaining more
realistic production ratios of heavy elements that might prove useful for the
U- and Th- chronologies.

In this paper we use a model with a proto-neutron star with (gravitational)
mass of $2.0 M_\odot$ and radius of 10 km, as in Wanajo et al. (2001), whose
nucleosynthesis results are in good agreement with the solar $r$-pattern for
nuclei $A \approx 120 - 200$.  We briefly describe the models here, and point
out some improvements that have been added.

The system is treated as time stationary and spherically symmetric, and the
radius of the neutron star is assumed to be the same as that of the neutrino
sphere. The physical variables in the neutrino wind are then functions of the
radius only. The ejected mass by neutrino heating is assumed to be negligible
compared to the mass of the neutron star. Therefore, the gravitational field in
which the neutrino-heated matter moves can be treated as a fixed-background
Schwarzschild metric.  Time variations of the temperature and density are then
solved by use of relations based on baryon, momentum, and energy conservation
(equations~(1)-(3) in Wanajo et al. 2001). The source term in the equation of
energy conservation is due to both heating and cooling by neutrino
interactions. The gravitational redshift of the neutrino energies, and the
bending of the neutrino trajectories expected from general relativistic
effects, are explicitly taken into account.

For the cooling by annihilation of electron-positron pairs into
neutrino-antineutrino pairs, the accurate table of Itoh et al. (1996) is
utilized. All other neutrino heating and cooling rates are taken from Otsuki et
al. (2000). The neutrino luminosities, $L_\nu$, of all neutrino flavors are
assumed to be equal, and the RMS average neutrino energies are taken to be 12,
22, and 34~MeV, for electron, anti-electron, and the other flavors of
neutrinos, respectively. The equation of state for the electron and positron
gas includes arbitrary relativistic pairs, which are of importance during
$\alpha$-processing. The electron fraction is set to be $Y_e
\approx 0.43$, which was mainly determined by the equilibrium of neutrino
capture on free nucleons with the above luminosities and energies of the
electron and anti-electron neutrinos at the neutrino sphere.

In this paper a {\it freezeout temperature}, $T_f$, is introduced to specify
the final temperature of the neutrino winds. As can be seen in realistic
hydrodynamic simulations of neutrino winds, all trajectories specific to the
$r$-process asymptotically approach the same temperature in the stalled shock
\citep{Woos94}. As discussed in \S~3, this assumption is also reasonable from
the nucleosynthetic point of view.  Otsuki et al. (2000) set the mass ejection
rate, $\dot M$, at the neutrino sphere of each wind so that the temperature at
10000~km in radius is 0.1~MeV. This assumption may be inadequate, as the
physical conditions of the neutrino sphere and the outer boundary are not
necessarily causally connected. On the other hand, Wanajo et al. (2001) set the
mass ejection rate to be $\dot M = 0.995 \times \dot M_c$, where $\dot M_c$ is
the maximum mass ejection rate at the neutrino sphere appropriate for the local
physical conditions (the wind with $\dot M_c$ becomes supersonic through the
sonic point, while those with $\dot M < \dot M_c$ are subsonic throughout).
However, the asymptotic temperatures differed significantly from wind to wind,
which is unlikely in a realistic situation.  In this paper, therefore, we have
specified both $\dot M$ and $T_f$ to avoid the above problems. The mass
ejection rate is set to be $\dot M = \dot M_c$. In fact, recent hydrodynamic
simulations show that the winds become supersonic, and then, in turn,
decelerate to become subsonic from the effects of the reverse shock coming from
the outer iron mantle \citep{Burr95}. $T_f$ is set to be the same for all
winds.  The temperature, $T$, of each wind is thus replaced with $T_f$ for all
$T < T_f$. The density is also replaced to that obtained at $T = T_f$. The
value of $T_f$ is determined in \S~3.

\section{THE $r$-PROCESS}

The $r$-process nucleosynthesis calculation, adopting the model described above
for the physical conditions, is obtained by application of an extensive
nuclear-reaction network code.  The network consists of $\sim 3600$ species,
all the way from single neutrons and protons up to the fermium isotopes ($Z =
100$, see Figure~2). We include all relevant reactions, i.e., $(n, \gamma)$,
$(p,\gamma)$, $(\alpha, \gamma)$, $(p, n)$, $(\alpha, p)$, $(\alpha, n)$, and
their inverses.  Reaction rates are taken from Thielemann (1995, private
communication) for nuclei with $Z
\le 46$ and from Cowan, Thielemann, \& Truran (1991) for those with $Z
\ge 47$. The latter used the mass formula by Hilf et al. (1976).

Our network does not include the nuclei near the neutron drip line for $Z <
10$, since no reliable data are yet available. Terasawa et al. (2001)
demonstrated that $r$-processing was hindered significantly by reaction flows
near the neutron drip line for $Z < 10$, owing to their rather low asymptotic
temperature ($= 6.2 \times 10^8$~K). As we suggest below, however, it may be
unlikely that the asymptotic temperature is, in reality, less than $\sim 8
\times 10^8$~K. Thus, inclusion of these reactions would not significantly
change our conclusions.

The three-body reaction $\alpha (\alpha n, \gamma)^9$Be, which is of special
importance as the bottleneck reaction to heavier nuclei, is taken from the
recent experimental data of Utsunomiya et al. (2001). The weak interactions,
such as $\beta$-decay, $\beta$-delayed neutron emission (up to three neutrons),
and electron capture are also included, although the latter is found to be
unimportant. Note that neutrino-induced reactions are not included. In fact,
neutrino capture on free nucleons might be of importance, as they may serve to
increase the electron fraction during seed element production
\citep{Meye98}, while those on heavy nuclei are likely to be unimportant.
However, it would be meaningless to consider these effects without taking the
accurate treatment of neutrino transport near the neutrino sphere into account.

The $\alpha$-decay chains and spontaneous fission processes are taken into
account only after the freezeout of all other reactions. For the latter, all
nuclei with $A \ge 256$ are assumed to decay by spontaneous fission only.  The
few known nuclei undergoing spontaneous fission for $A < 256$ are also
included, along with their branching ratios. Neutron-induced and
$\beta$-delayed fissions, as well as the contribution of fission fragments to
the lighter nuclei, are neglected. Obviously, these treatments of the fission
reactions are oversimplified. Nevertheless, this may be acceptable, at least to
first order, as we demonstrate below. We leave the more accurate treatment of
these matters to future work. In particular, $\beta$-delayed fission may be of
importance, as shown by Schatz et al. (2002).

Each calculation is started at $T_9 = 9$ (where $T_9 \equiv T/10^9$~K). At this
point the nuclear statistical equilibrium (NSE) consists mostly of free
nucleons. The initial mass fractions of neutrons and protons are thus given by
$X_n = 1 - Y_e$ and $X_p = Y_e$, respectively. In Wanajo et al.  (2001), the
initial electron fraction $Y_e$ was fixed to be 0.40.  In this paper $Y_e$ is
varied from 0.39 to 0.49 to examine its effect on the $r$-process. Note that,
for simplicity, for each $Y_e$ the same trajectories are used, being derived
with $Y_e \approx 0.43$.  The small change of $Y_e$ is, however, found to be
unimportant for the temperature and density histories.

The freezeout temperature, $T_f$, is determined as follows. At present, there
is no positive evidence that the $r$-process pattern between the
$r$-process-enhanced stars (e.g., CS~31082-001) and the solar abundances
differs for $Z \ge 56$, except for Th and U \citep{Hill02}. Thus, since our
main goal is to estimate reliable yields of U and Th, the local $r$-process
pattern near the third peak ($^{195}$Pt) is treated most carefully, in order to
reproduce the solar $r$-process pattern. In Figure~1 the nucleosynthesis
results before and after $\alpha$-decay and fission (thick line), obtained with
different $T_{9f} (\equiv T_f/10^9$~K), are compared to the solar $r$-process
pattern (points; K\"appeler et al. 1989) normalized to the height of the
third-peak nuclei. $L_\nu$ and $Y_e$ are taken to be $10^{51.8}$~ergs~s$^{-1}$
and $0.43$, respectively. The thin line denotes the results obtained by turning
off the process of $\beta$-delayed neutron emission. Note that the local
pattern of the $r$-element abundances (i.e., near $A = 195$) is not
significantly altered as the result of changing $L_\nu$ and $Y_e$ (see
Figures~4 and 5). Thus, the discussion below may hold even for superposition of
the yields from multiple trajectories.

For $T_{9f} = 0.6 - 0.8$ (Figures~1a-c), the position of the third peak shifts
to the low side by two units from $A = 195$.  Moreover, the valley on the lower
side of the third peak seen in the solar $r$-process pattern ($A
\approx 183$) is substantially filled up with the jagged pattern. On the other
hand, for $T_{9f} = 1.2 - 1.4$ (Figures~1g-i), the third peak significantly
shifts to the high side from $A = 195$. In addition, the abundance pattern is
unacceptably jagged compared to the solar $r$-process pattern. In this regard,
the results with $T_{9f} = 0.9 - 1.1$ reproduce the properties of the solar
$r$-pattern near the third peak quite well (Figures~1d-f).  Indeed, the third
peak is correctly located at $A = 195$. The pattern in the valley at $A
\approx 183$ is also reasonably reproduced.  In this paper, therefore, we set
$T_{9f}$ to be 1.0, so that the predicted $r$-process pattern is in the best
agreement with that of the solar $r$-process pattern near the third peak. Note
that the second peak ($A = 130$) nuclei in Figure~1 are substantially
underproduced compared to the solar $r$-abundances in this specific trajectory,
which originates mainly from the higher $L_\nu$ trajectories (Figures~4 and 5).

In Table~1, the abundance ratios among Eu, Ir, Pb, Bi, Th, and U are listed
along with the isotope number fraction of $^{151}$Eu. Note that, in this table,
``Pb'' means the sum of $^{206}$Pb, $^{207}$Pb, $^{208}$Pb, and $^{235}$U,
while ``U'' means $^{238}$U, owing to the short mean half-life of $^{235}$U
compared to the cosmic age. As can be seen, the ratio U/Th is insensitive to
the change of $T_{9f}$, except for the higher temperature cases with $T_{9f}
\ge 1.2$. This is simply due to the fact that Th and U are close to one another
in mass number. Thus, small variations of $T_{9f}$ from 1.0 may not change the
chronometric ages substantially, as long as the U-Th pair is employed. In
contrast, other abundance ratios are found to be quite sensitive to the change
of $T_{9f}$. For example, the ratio Th/Eu varies by more than a factor of two
in the same temperature range. This implies that the chronometer pairs Th-$r$
may be unreliable unless the astrophysical model is strictly constrained. It is
interesting to note that the ratio between Pb-Bi and Th-U can be changed by a
factor of two even though they are mainly $\alpha$-decay products.

The isotope ratio of Eu can provide another constraint on the determination of
$T_{9f}$.  Recent spectroscopic studies have shown that the isotope number
fractions of a few extremely metal-poor stars, including CS~22892-052
\citep{Sned02} and CS~31082-001 (Aoki et al., in preparation) are close to
the solar value of 0.478. This is also in good agreement with the calculated
values for $T_{9f} \lesssim 1.0$ in the last column of Table~1.

The strong influence of $T_f$ on the final predicted $r$-process pattern
can be recognized as the consequence of the freezeout effects. Figure~2
shows the post-freezeout evolution of the $r$-process abundances for
$T_{9f} = 1.0$ with $L_\nu = 10^{51.8}$~ergs~s$^{-1}$ and $Y_e =
0.43$. The abundances are color-coded with their pattern as a function
of mass number in the upper left.  The dots denote the nuclei included
in the reaction network, with the stable and long-lived nuclei shown as
larger dots.  At the freezeout of $r$-processing, which occurs 0.48
seconds after $T_9 = 9$, a clear odd-even effect appears, owing to the
$(n, \gamma) - (\gamma, n)$ equilibrium (Figure~2a). After 0.58 seconds,
the abundances near the third peak are still along the neutron magic
number $N = 126$ (Figure~2b). As a result, the third peak slightly
shifts to the larger side in mass number, due to the small fraction of
neutrons emitted by photo-dissociation of fragile nuclei (e.g., $A \sim
183$) and by the $\beta$-delayed process; these are quickly absorbed by
other nuclei (including those near the third peak). The abundance valley
at $A \approx 183$ is then sharpened.  At the same time the abundance
pattern is significantly smoothed. At some point the abundances near
the third peak separate from $N = 126$ due to $\beta$-decay, becoming
more and more robust against photo-dissociation, as well as with the
decrease in the $\beta$-delayed process (Figure~2c). The position of the
third peak is thus fixed. The abundance pattern is smoothed further by
the continuation of non-equilibrium neutron emission and absorption
among fragile nuclei. It is important to note that the separation from
the $N = 126$ line takes place at earlier times for lower $T_f$, and
vice versa, which determines the location of the third peak, as well as
the smoothness of the pattern near it.  Although the $\beta$-delayed
process is found to be unimportant for setting the locations of the
second ($A \approx 130$) and the third ($A \approx 195$) peaks, it does
contribute to the enhancement of nuclei with $A > 220$, in particular
for $T_{9f} \ge 0.9$, as can be seen in Figure~1. This results in the
enhancement of Th-U/Pb-Bi for $T_{9f} \sim 1.0$ found in Table~1.

In order to calculate the mass-integrated $r$-process yields, the time
evolution of the neutrino luminosity, $L_\nu$, is assumed to be $L_\nu (t) =
L_{\nu 0} (t/t_0)^{-1}$.  $L_{\nu 0}$ and $t_0$ are taken to be
$10^{52.6}$~ergs~s$^{-1}$ ($\approx 4 \times 10^{52}$~ergs~s$^{-1}$) and 0.2~s,
respectively, being in reasonable agreement with the hydrodynamic results of
Woosley et al. (1994). The time evolution of the neutrino sphere radius is also
assumed to be $R_\nu (t) = R_{\nu 0} (t/t_0)^{-1/3}$, where $R_{\nu 0} =
15$~km. $R_\nu$ is replaced with 10~km for $R_\nu < 10$~km. This rather rapid
contraction of $R_\nu$ is required for robust $r$-processing to occur at early
times with a compact proto-neutron star model \citep{Wana01, Thom01}. Figure~3
shows $R_\nu$ (km) and $\dot M_c$ ($M_\odot$~s$^{-1}$), as well as the entropy
per baryon $S$ ($N_A k$) and the timescale of seed abundance production $\tau$
(msec), as functions of $L_\nu$. The timescale, $\tau$, is defined as the
duration from 6 to 2.5 in $T_9$, which approximately corresponds to the phase
of $\alpha$-processing, i.e., the seed abundance production prior to
$r$-processing. Note that our definition of $\tau$ differs from the $e$-folding
time of the density often taken in other work, since the latter is not
necessarily a good indicator of the seed abundance production. For $L_\nu
\approx 5 - 10$~ergs~s$^{-1}$, $\tau$ is significantly short ($\approx
5-10$~msec) and $S$ is moderately high ($\approx 140-160 N_A k$), which are
suitable conditions for the $r$-process (preventing consumption of free
neutrons prior to $r$-processing). For $L_\nu \gtrsim 10^{52}$~ergs~s$^{-1}$,
$\tau$ increases and $S$ decreases rapidly with $L_\nu$, owing to the
increasing $R_\nu$. The total ejected mass, calculated by following the time
evolution of $L_\nu$ and $\dot M_c$, is $5.0 \times 10^{-4} M_\odot$, more than
one order of magnitude smaller than that of Woosley et al. (1994). This is
perhaps due to the fact that our assumption of the steady-state flow does not
hold well for $L_\nu > 10^{52}$~ergs~s$^{-1}$. This high luminosity phase is,
however, unimportant for the $r$-process (Figure~4).

The nucleosynthesis results for different $L_\nu$ with $Y_e = 0.43$, before
$\alpha$-decay and spontaneous fission, are shown in Figure~4. For $\log L_\nu
({\rm ergs~s}^{-1}) = 52.6$, 52.4, and 52.2, no significant $r$-processing
beyond the second peak occurs (Figures~4a-c). This is due to the low $S$ ($=
61, 81, 110 N_A k$), regardless of the short $\tau$ (= 8.9, 7.3, 5.9~msec). For
$\log L_\nu ({\rm ergs~s}^{-1}) = 52.0$, 51.8, and 51.6, robust $r$-processing
occurs, producing the second- and third-peak nuclei, and substantial actinides
(Figure~4d-f). This is a consequence of the rather short timescale of seed
abundance production ($\tau = 5.5$, 7.5, 11 msec) and the high entropy per
baryon ($S = 141$, 154, and 169 $N_A k$). For $\log L_\nu ({\rm ergs~s}^{-1}) =
51.4$, 51.2, and 51.0, the second and the third peaks are still prominent, but
production of the actinides is not of importance with high $S$ (= 184, 200, 217
$N_A k$) and long $\tau$ (= 17, 26, and 41~msec, Figure~4g-h). Note that, for
$\log L_\nu ({\rm ergs~s}^{-1}) = 52.0-51.0$, the positions of the second ($A
\approx 130$) and third ($A \approx 195$) peaks and the valley at $A \approx
183$ do not vary significantly (nor do the local patterns close to these
features), although the global patterns are different (Figures~4d-i). This is
due to the use of the same $T_f$ ($= 1.0 \times 10^9$~K) in all situations.

We now integrate the nucleosynthesis results of trajectories with $\log
L_\nu ({\rm ergs~s}^{-1}) = 52.6-51.0$, using the time evolution of
$L_\nu$ with $\dot M_c$ (Figure~3). The interval of $L_\nu$ is taken to
be 0.1~dex. These $L_\nu$ correspond to times $t = 0.2-8.0$~sec at the
neutrino sphere. The abundances of species formed after this period are
neglected because of the small $\dot M_c$ ($< 2 \times 10^{-6}
M_\odot$~s$^{-1}$). In Figure~5, the mass-integrated abundances are
compared with the solar $r$-process pattern \citep{Kapp89} scaled to
match $^{153}$Eu. The $\alpha$-decay and spontaneous fission after
freezeout of all other reactions are included. The initial electron
fraction, $Y_e$, is varied from 0.39 to 0.49 to demonstrate its effect
on the $r$-process.  Note that in our models $Y_e$ is taken to be
constant with time.  Although $Y_e$ should certainly be a function of
time, it is meaningless to consider this refinement without taking into
account accurate treatment of neutrino transport near the neutrino
sphere. As can be seen in Figure~5, for $Y_e = 0.39-0.43$, the calculated
abundances are in reasonable agreement with the solar $r$-process pattern
between $A \approx 120$ and 200. For larger neutron excess, with $Y_e =
0.39-0.41$, $^{232}$Th and $^{235, 238}$U are significantly overproduced
compared to the solar $r$-process pattern (Figures~5a-c), as are other (stable)
nuclei near the third peak. For $Y_e = 0.42$, overall agreement of the
abundance pattern of $A \ge 120$ to the solar $r$-process pattern can be seen
(Figure~5d). For $Y_e = 0.43$, Th and U are significantly underproduced
compared to the solar $r$-process pattern (Figure~5e). In the current models,
therefore, $Y_e$ should be $\le 0.42$ to achieve sufficient production of
Th and U. Note that the location of the third ($A \approx 195$) peak and the
valley at $A \approx 183$ still hold even after integration of the yields. The
deficiency of nuclei at $A = 130-140$, seen for all $Y_e$ cases, could be due
to our insufficient treatment of the physics of the neutrino winds for $L_\nu >
10^{52}$~ergs~s$^{-1}$, in which the assumption of the steady state winds is
not manifest at all. More likely, this might be due to the nuclear properties
of the mass model \citep{Hilf76} employed in this study, as shown by
Freiburghaus et al. (1999a). In their paper, calculations with other mass
formulae, FRDM \citep{Moel95} and ETFSI \citep{Abou95}, were able to well
reproduce the solar $r$-abundances in this mass region.

As can be seen in Figure~5, overproduction of nuclei with $A \approx 90-110$ is
evident, except for the $Y_e = 0.49$ case. This is a consequence of the large
$\dot M_c$ for $L_\nu > 10^{52}$~ergs~s$^{-1}$ (Figure~3), in which little
$r$-processing occurs (Figures~4a-c). For high $Y_e$ (= 0.49, Figure~5f), the
overproduction significantly diminishes. This arises because, in the nuclide
chart (see Figure~2), $Y_{e,\, {\rm seed}} > 0.446$ lines have intersections
with $\alpha$ separation energies of $\approx 6$~MeV on the lower side of the
neutron magic number $N = 50$ \citep{Hoff96, Frei99a}. At the same time,
however, the $r$-process abundance of $A \approx 120-200$ decreases
substantially. This is due to the small levels of the seed abundance, owing to
the extremely strong $\alpha$-rich freezeout by the less efficient $\alpha
(\alpha n, \gamma)^9$Be reaction with low neutron excess \citep{Hoff97}.

In Table~2, the total ejected masses of $r$-process abundances ($A \ge
120$) are shown, along with that of Eu. For $Y_e = 0.49$, the ejected
mass is about one order of magnitude smaller than for other $Y_e$ cases. 
According to the chemical evolution study of the Galactic halo by
Ishimaru \& Wanajo (1999), $\sim 5 \times 10^{-7} M_\odot$ of Eu per
$r$-process event is needed to reproduce the large dispersion of
observed Eu/Fe in extremely metal-poor stars. The mass of Eu obtained
for $Y_e = 0.49$ is too small compared to that level, while remaining in
reasonable agreement for lower $Y_e$ cases. We speculate that, if the
primary $r$-process site is {\it really} neutrino winds, as discussed in
this study, the electron fraction changes from $Y_e \approx 0.49$ to
$\lesssim 0.42$ rapidly with time.  In this case, the overproduction would not
appear at the early phase ($L_\nu > 10^{52}$~ergs~s$^{-1}$), but would be
followed by later $r$-nuclei ($A > 120$) production ($L_\nu <
10^{52}$~ergs~s$^{-1}$). As can be seen in the third column in Table~2, except
for $Y_e = 0.49$, the isotopic fraction of $^{151}$Eu is in good agreement with
both the solar value and the observational estimates (Sneden et al. 2002; Aoki
et al., in preparation), even after mass-integration of the yields.

The last column of Table~2 shows the mass of fission fragments, whose
contribution to the lighter nuclei is neglected in this study. This assumption
appears adequate since the contribution of fission fragments is less than $5
\%$ of the total mass of nuclei with $A\ge 120$. This follows since, in this
study, the neutron-to-seed ratio at the onset of $r$-processing ($T_9 = 2.5$)
is 138 at most, with $\langle A_{\rm seed} \rangle \approx 95$. This results in
no significant production of nuclei with $A \ge 255$, where fission dominates.

\section{THE U-Th COSMOCHRONOLOGY}

In Figure~6 the available spectroscopic abundance data for CS~31082-001 (dots)
are compared with the nucleosynthesis results discussed in \S~3 (thick line)
and with the solar $r$-pattern (thin line), scaled at Eu ($Z = 63$). The data
for the neutron-capture elements in this star are taken from Hill et al.
(2002).  CS~31082-001 is the first extremely metal-poor star ($[{\rm Fe/H}] =
-2.9$) in which the UII line (3859.59 \AA) has been detected \citep{Cayr01}.
The abundance pattern of this star between Ba ($Z = 56$) and Ir ($Z = 77$) is
in excellent agreement with the solar $r$-process pattern, except for a few
elements (Tb, Hf, Os).  It is possible that, in these few cases, the
disagreements are due to difficulties in the abundance estimates
\citep{Hill02}. Note that some lighter elements (Rh, Ag) are not in agreement
with the solar $r$-process pattern, as has also been seen in CS~22892-052
\citep{Sned00}, and which implies the presence of two distinct
$r$-process sites \citep{Ishi00,Qian00}.

Nucleosynthesis results for all $Y_e$ cases are in good agreement with the
abundance pattern of CS~31082-001 for the elements between Pr ($Z = 59$) and Tm
($Z = 69$), with the exception of Tb. Note that a few of the predicted
abundances for elements near the second peak (Ba, La, and Ce) are significantly
deficient compared to both the observational and solar patterns. As mentioned
in \S~3, this might be due to the properties of the nuclear mass model employed
in this study.  In future work we plan to check if these elements are
reproduced with other mass formulae, since the observational data suggest the
universality of the $r$-process pattern between the second and the third
$r$-process peaks. The platinum-peak elements Os and Ir are in reasonable
agreement, except for the $Y_e = 0.49$ case, although the height of the
$^{195}$Pt differs. In addition, the large abundance of Os in CS~31082-001
compared to the scaled solar $r$-process pattern makes it questionable if the
third peak is really located at Pt. It would clearly be useful to obtain
measurements of the Pt abundance in this star, so that the height, as well as
the position, of the third peak could be better constrained.  Note that
$^{235}$U is assumed to have $\alpha$-decayed away because of its relatively
short half life (0.704 Gyr).

As discussed by Hill et al. (2002), the age of CS~31082-001 can be inferred by
application of one or more of the following three relations:

\begin{eqnarray}
t^*_{{\rm Th}, r} & = & 
46.67\, [\log ({\rm Th}/r)_0 - \log ({\rm Th}/r)_{\rm now}]\ {\rm Gyr}
\\
t^*_{{\rm U},r} & = & 
14.84\, [\log ({\rm U}/r)_0 - \log ({\rm U}/r)_{\rm now}]\ {\rm Gyr}
\\
t^*_{{\rm U, Th}} & = & 
21.76\, [\log ({\rm U/Th})_0 - \log ({\rm U/Th})_{\rm now}]\ {\rm Gyr}
\end{eqnarray}

\noindent with the half lives of $^{232}$Th (14.05 Gyr) and $^{238}$U
(4.468 Gyr), $r$ is a stable $r$-element, and the subscripts ``0'' and
``now'' denote the initial and current values derived by theory and
observation, respectively. In principle, $t^*_{{\rm U},r}$ and
$t^*_{{\rm U, Th}}$ can be more precise chronometers than $t^*_{{\rm
Th}, r}$, owing to the smaller coefficients in front of equations~(2)
and (3). In practice, the U-Th pair (equation~(3)) is expected to be the
most precise chronometer owing to the similar atomic properties of these
species, which serve to diminish the observational errors associated
with determination of their ratio. Furthermore, the ratio U/Th is less
dependant on the theoretical model used, since these species are
separated by only two units in atomic number, as well as being mostly
the result of $\alpha$-decayed products (Table~1). As stable elements
$r$, Eu, Os, and Ir are selected. The Eu abundance derived by
spectroscopic analyses is more reliable than for any other elements with
$Z \ge 59$. The Pt-peak elements Os and Ir are co-produced with Th and U
in the same (or close in $L_\nu$) trajectories, and thus are perhaps
less model dependent as well.  Note that, if possible, the Pb abundance
in this star should be used, since it is mainly synthesized by the same
$\alpha$-decay chains as Th and U.  However, Pb has not yet been
detected in CS~31082-001, and for now only an upper limit to its
abundance is available.  For reasons that are not yet clear, the upper
bound on Pb reported by Hill et al. (2002) is lower than the solar
value, being satisfied only by the $Y_e \le 0.42$ cases in which Th and
U are not significantly produced.  Thus, Pb is omitted for the age
determination in this study, although its future measurement would
clearly be of importance.

Table~3 lists the predicted initial abundance ratios among Eu, Os, Ir,
Pt, Th, and U ($A = 238$) for all the $Y_e$ cases considered in the
present study. The ratio U/Th is found to be insensitive to $Y_e$
(except for the extreme case of $Y_e = 0.49$), which is in excellent
agreement with those constrained by the solar $r$-process abundances in
\citet{Gori01}.  It should be noted that these values can be applied not
only to CS~31082-001, but also to other metal-poor halo stars in which
uranium and thorium (and platinum) are detected by future observations
\citep{Cowa02}. With the use of these ratios as the initial values, the
age of CS~31082-001 is derived as shown in Table~4, as well as in
Figure~7. It is noteworthy that the age obtained from the U-Th
chronometer pair is considerably more robust than the alternatives,
resulting in a range $\sim 13.5-14.2$~Gyr, with an exception of 12.2~Gyr
for the $Y_e = 0.49$ case.  Therefore, we use only the U-Th chronometer
for dating of this star, while the others are regarded as consistency
checks only. In particular, the Th-Eu pair should be used to constrain
the model parameter $Y_e$ in this study, the species being widely
separated in mass number. For this reason, $Y_e$ is constrained to be
$\approx 0.40$, owing to the consistency between ages based on the U-Th
pair and others (especially Th-Eu). The ages obtained by use of Os are
significantly higher than those by others, because of the previously
noted discrepancies between observation and the theoretical prediction
(Figure~6). As seen in Table~4, the difference between the ages obtained from
U-Th between $Y_e = 0.39$ and 0.41 is only 0.1~Gyr. On the other hand, the
observational error associated with the U/Th ratio, 0.11 dex, leads to an
uncertainty of 2.4~Gyr from application of equation~(3).  We conclude,
therefore, that the age of CS~31082-001 is $14.1 \pm 2.5$~Gyr.

It should be emphasized that the age determination made in the present paper is
derived from the exploration of a single specific model based on the
neutrino-wind scenario, and relies as well on the choice of one of the many
existing nuclear data sets. Thus, it is clear that our final result may be
changed if we were to make use of other astrophysical models. The nuclear mass
models can also significantly affect the age determination
\citep{Gori01}. Obviously, more study with other astrophysical models, as
well as with a number of different nuclear data sets, are needed to
derive the final results.

\section{SUMMARY AND CONCLUSIONS}

We have constructed a robust model of the $r$-process based on the
neutrino-wind scenario for application to the U-Th cosmochronology. The model
of neutrino winds was taken from Otsuki et al. (2000) and Wanajo et al. (2001),
along with some improvements described above. The $r$-process with various
neutron star masses ($M_{\rm NS}$) and radii ($\approx R_\nu$) has already been
explored extensively by Wanajo et al.  (2001). Thus, $M_{\rm NS}$ and
(asymptotic) $R_\nu$ were fixed to be $2.0 M_\odot$ and 10~km, respectively,
for which the most robust $r$-processing was realized. On the other hand, the
initial electron fraction $Y_e$ (assumed constant with time) was taken to be a
free parameter, rather than fixed to be 0.40 as in Wanajo et al. (2001).

The model was tuned to fit the solar $r$-process pattern, since recent
spectroscopic studies of extremely metal-poor stars have indicated a {\it
universal} (solar $r$-like) abundance pattern, at least between the second and
third r-process peaks.  Attention was paid to reproduce the pattern near $A
\approx 195$, being close to Th and U in mass number. We introduced the
freezeout temperature, $T_f$, as the most crucial parameter for constructing
the model.  Comparing the nucleosynthesis results with the solar $r$-process
pattern, $T_f$ was determined to be $1.0 \times 10^9$~K. The nucleosynthesis
yields were mass-integrated assuming simple evolutions of $L_\nu$ and $R_\nu$.
The results between $A \approx 120$ and 200 for $Y_e = 0.39-0.43$ models were
in good agreement with the solar $r$-pattern, although the production of Th and
U differed from model to model.

The results of our nucleosynthesis predictions were compared to the
spectroscopic abundance pattern of the $r$-process-enhanced metal-poor star
CS~31082-001.  Explorations of various combinations between the actinides and
stable isotopes indicated that the $Y_e = 0.40$ case was found to be the best
models for the age determination in this study. As a result, the age of
CS~31082-001 was determined to be $14.1 \pm 2.5$~Gyr by use of the U-Th
chronometer, which can be regarded as a hard lower limit on the age of the
universe. In fact, this age is in excellent agreement with the age of the
universe derived by other dating techniques (e.g., globular clusters, Type Ia
supernovae, and the cosmic-microwave background), as well as with other stellar
radioactive age estimates using the Th-Eu \citep{Sned96, Cowa97} and U-Th
\citep{Hill02, Scha02, Cowa02} pairs.  Such a strong constraint on the
age of this star was only made possible by the use of U and Th in conjunction
with one another.  Without uranium, the chronometers involving Th alone
presented only rather loose constraints.

In this study the Pb abundance in CS~31082-001 was not available for
age determination, since only an upper limit on its abundance has been
presented to date. It should be noted that the predicted Pb abundance
for $Y_e =0.40$ case is {\it higher} than this upper limit, which seems
difficult to solve \citep{Scha02}. It is noteworthy, however, that
freezeout effects can increase the ratio Th-U/Pb-Bi (see \S~3), which should be
examined further, as well as with other nuclear data sets. It remains possible
that the solar $r$-process Pb abundance, inferred from the residual of the solar
$s$-process components, is not correct \citep{Arla99}. Clearly,
re-investigation of the Pb abundance in CS~31082-001, as well expansion of the
searches for additional $r$-process-enhanced metal-poor stars, in which both U
and Th are measurable, are highly desirable in the near future.  It is also of
importance to obtain measurements of the Pt abundance in CS~31082-001 (which is
only possible with the strong Pt absorption features observable in the near UV
from space), so as to constrain the height, as well as the position, of the
third peak \citep{Cowa02}.

We emphasize that the conclusions in this study should be considered as merely
the first step toward a better understanding of the dating technique involving
the U-Th chronometer. In particular, we note that the inferred age of
CS~31082-001 could be significantly changed if we had utilized a different
nuclear mass model, as demonstrated by Griely and Arnould (2001). We expect,
however, that a detailed comparison of the nucleosynthesis results with the
solar $r$-process pattern, as in \S~3, can distinguish better mass models. We
suggest that the freezeout temperature, $T_f$ (or the asymptotic temperature in
more realistic models), is a useful parameter, as the freezeout effects are of
importance to the final $r$-process pattern. While the waiting-point
approximation, often referred to as the ``classical $r$-process model,'' has
been an effective (and thus far, the only) method used to test the nuclear mass
models, we should keep in mind that non-equilibrium processes can lead to
non-negligible effects, as discussed in \S~3 (e.g., the positions of the
$r$-process peaks, the smoothness of abundance pattern, and the ratio
Th-U/Pb-Bi).

It should also be noted that we would not insist that the neutrino-wind model
explored in this study is necessarily the best (or only) astrophysical site for
the $r$-process. In fact, the physical conditions in this model, namely, the
rather massive proto-neutron star ($2.0 M_\odot$) with rapid contraction of
$R_\nu$ to be 10~km, seem difficult to be achieved according to the current
hydrodynamic studies. Clearly, more study is needed to seek the true
$r$-process site. Nevertheless, it is encouraging that the age of the
CS~31082-001 determined in this study is in excellent agreement with the age of
the universe derived by other dating techniques. We hope our results provide
some fresh insights to the future modeling of the true astrophysical
$r$-process site.

\acknowledgments

We would like to acknowledge V. Hill for providing an up-to-date
abundance table for CS~31082-001 prior to publication. We also thank
H. Schatz for helpful discussions, and the referee, J. J. Cowan, who
helped us to improve this paper. S. W. thanks K. Sumiyoshi and H.
Utsunomiya for providing data for the $\alpha (\alpha, n)^9$Be reaction.
This work was supported by a Grant-in-Aid for Scientific Research
(13740129) from the Ministry of Education, Culture, Sports, Science, and
Technology of Japan.  T.C.B acknowledges partial support from NSF grants
AST-00 98549 and AST-00 98508.  T.C.B. would also like to recognize
partial support from an international scholar fellowship from the
Ministry of Education, Culture, Sports, Science, and Technology of
Japan, which supported his sabbatical stay in Tokyo, where discussions
leading to the present paper were initiated.

\clearpage

\begin{deluxetable}{ccccccccccc}
%\footnotesize
\tablecaption{Abundance Ratios \label{tab1}}
\tablewidth{0pt}
\tablehead{
\colhead{$T_{9f}$} & \colhead{Th/Eu} & \colhead{Th/Ir} & \colhead{Th/Pb\tablenotemark{a}} & \colhead{Th/Bi} & \colhead{U/Eu\tablenotemark{a}} & \colhead{U/Ir\tablenotemark{a}} & \colhead{U/Pb\tablenotemark{a}} & \colhead{U/Bi\tablenotemark{a}} & \colhead{U/Th\tablenotemark{a}} & \colhead{$f$($^{151}$Eu)\tablenotemark{b}}}
\startdata
   0.60 &    0.24 &    0.02 &    0.08 &    0.21 &    0.11 &    0.01 &    0.04 &    0.10 &    0.44 &    0.47 \\
   0.70 &    0.36 &    0.02 &    0.09 &    0.27 &    0.17 &    0.01 &    0.04 &    0.13 &    0.47 &    0.46 \\
   0.80 &    0.55 &    0.02 &    0.13 &    0.36 &    0.25 &    0.01 &    0.06 &    0.17 &    0.46 &    0.45 \\
   0.90 &    0.46 &    0.02 &    0.14 &    0.50 &    0.23 &    0.01 &    0.07 &    0.25 &    0.50 &    0.44 \\
   1.00 &    0.38 &    0.02 &    0.15 &    0.45 &    0.19 &    0.01 &    0.07 &    0.22 &    0.49 &    0.52 \\
   1.10 &    0.41 &    0.03 &    0.14 &    0.47 &    0.18 &    0.02 &    0.06 &    0.21 &    0.44 &    0.41 \\
   1.20 &    0.17 &    0.06 &    0.09 &    0.46 &    0.06 &    0.02 &    0.03 &    0.16 &    0.36 &    0.40 \\
   1.30 &    0.05 &    0.09 &    0.05 &    0.23 &    0.01 &    0.03 &    0.02 &    0.07 &    0.32 &    0.24 \\
   1.40 &    0.03 &    0.15 &    0.06 &    0.27 &    0.01 &    0.03 &    0.01 &    0.05 &    0.18 &    0.19
\enddata
\tablenotetext{a}{Pb = $^{206}$Pb + $^{207}$Pb + $^{208}$Pb + $^{235}$U, 
                  U  = $^{238}$U (see text)}
\tablenotetext{b}{isotope number fraction of $^{151}$Eu}
\end{deluxetable}

%\clearpage

\begin{deluxetable}{ccccc}
%\footnotesize
\tablecaption{Ejected Mass ($M_\odot$) \label{tab2}}
\tablewidth{0pt}
\tablehead{
\colhead{$Y_e$} & \colhead{$A \ge 120$} & \colhead{Eu} &
 \colhead{$f$($^{151}$Eu)\tablenotemark{a}} & \colhead{fission\tablenotemark{b}}}
\startdata
   0.39 &   $1.5\times 10^{-4}$ &   $6.2\times 10^{-7}$ &    0.50 &   0.042 \\
   0.40 &   $1.2\times 10^{-4}$ &   $5.4\times 10^{-7}$ &    0.50 &   0.025 \\
   0.41 &   $9.5\times 10^{-5}$ &   $4.3\times 10^{-7}$ &    0.52 &   0.014 \\
   0.42 &   $7.4\times 10^{-5}$ &   $4.5\times 10^{-7}$ &    0.51 &   0.005 \\
   0.43 &   $5.6\times 10^{-5}$ &   $4.6\times 10^{-7}$ &    0.51 &   0.002 \\
   0.49 &   $4.5\times 10^{-6}$ &   $6.8\times 10^{-8}$ &    0.40 &   0.000
\enddata
\tablenotetext{a}{isotope number fraction of $^{151}$Eu}
\tablenotetext{b}{mass fraction of fission fragments ($A \ge 120$)}
\end{deluxetable}

%\clearpage

\begin{deluxetable}{rrrrrrrrrr}
%\footnotesize
\tablecaption{Abundance Ratio (Log Scale) \label{tab3}}
\tablewidth{0pt}
\tablehead{
\colhead{$Y_e$} &
\colhead{Th/Eu} & \colhead{Th/Os} & \colhead{Th/Ir} & \colhead{Th/Pt} &
\colhead{U/Eu} & \colhead{U/Os} & \colhead{U/Ir} & \colhead{U/Pt} &
\colhead{U/Th}}
\startdata
   0.39 &    0.18 &   $-$0.18 &   $-$0.60 &   $-$1.08 &   $-$0.11 &   $-$0.47 &   $-$0.89 &   $-$1.37 &   $-$0.29 \\
   0.40 &    0.05 &   $-$0.41 &   $-$0.84 &   $-$1.27 &   $-$0.24 &   $-$0.70 &   $-$1.13 &   $-$1.57 &   $-$0.29 \\
   0.41 &   $-$0.11 &   $-$0.68 &   $-$1.10 &   $-$1.48 &   $-$0.40 &   $-$0.97 &   $-$1.39 &   $-$1.76 &   $-$0.29 \\
   0.42 &   $-$0.58 &   $-$1.16 &   $-$1.56 &   $-$1.85 &   $-$0.89 &   $-$1.47 &   $-$1.87 &   $-$2.16 &   $-$0.31 \\
   0.43 &   $-$0.92 &   $-$1.43 &   $-$1.82 &   $-$2.06 &   $-$1.24 &   $-$1.75 &   $-$2.13 &   $-$2.38 &   $-$0.32 \\
   0.49 &   $-$2.75 &   $-$2.50 &   $-$2.83 &   $-$3.17 &   $-$3.13 &   $-$2.89 &   $-$3.21 &   $-$3.55 &   $-$0.38
\enddata
\end{deluxetable}

%\clearpage

\begin{deluxetable}{rrrrrrrr}
%\footnotesize
\tablecaption{Ages of CS~31082-001 (Gyr) \label{tab4}}
\tablewidth{0pt}
\tablehead{
\colhead{$Y_e$} &
\colhead{Th/Eu} & \colhead{Th/Os} &
\colhead{Th/Ir} & \colhead{U/Eu} &
\colhead{U/Os} & \colhead{U/Ir} &
\colhead{U/Th}}
\startdata
   0.39 &   18.77 &   57.52 &   27.18 &   15.63 &   27.95 &   18.30 &   14.16 \\
   0.40 &   12.61 &   46.73 &   16.09 &   13.62 &   24.47 &   14.73 &   14.10 \\
   0.41 &    5.17 &   34.01 &    3.64 &   11.32 &   20.49 &   10.83 &   14.19 \\
   0.42 &  $-$16.90 &   11.67 &  $-$17.55 &    3.97 &   13.05 &    3.76 &   13.70 \\
   0.43 &  $-$32.54 &   $-$0.84 &  $-$29.64 &   $-$1.12 &    8.96 &   $-$0.20 &   13.53 \\
   0.49 & $-$118.21 &  $-$51.05 &  $-$76.97 &  $-$29.30 &   $-$7.94 &  $-$16.18 &   12.16
\enddata
\end{deluxetable}

\clearpage
\begin{figure*}
\epsscale{2.0}
\plotone{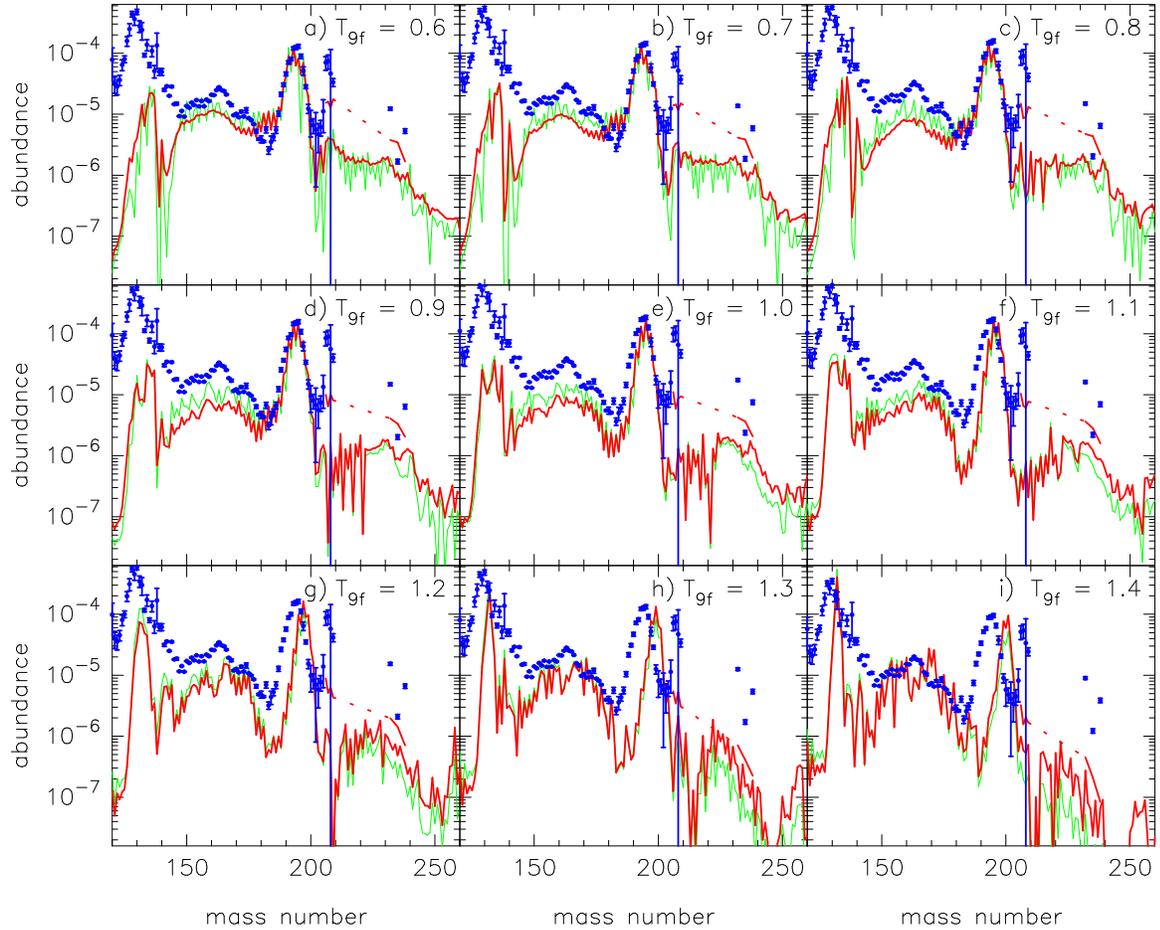}
\caption{Comparison of the nucleosynthesis yields (line) for $T_f = $
(a) 0.6, (b) 0.7, (c) 0.8, (d) 0.9, (e) 1.0, (f) 1.1, (g) 1.2, (h) 1.3,
and (i) 1.4, with the solar $r$-pattern (dots), scaled at the height of
the third peak for $\log L_\nu$~(ergs~s$^{-1}) = 51.8$ with $Y_e = 0.43$.
The thick and thin lines represent situations with and without
$\beta$-delayed neutron emission, respectively.}
\end{figure*}

\clearpage
\begin{figure*}
\epsscale{1.5}
\plotone{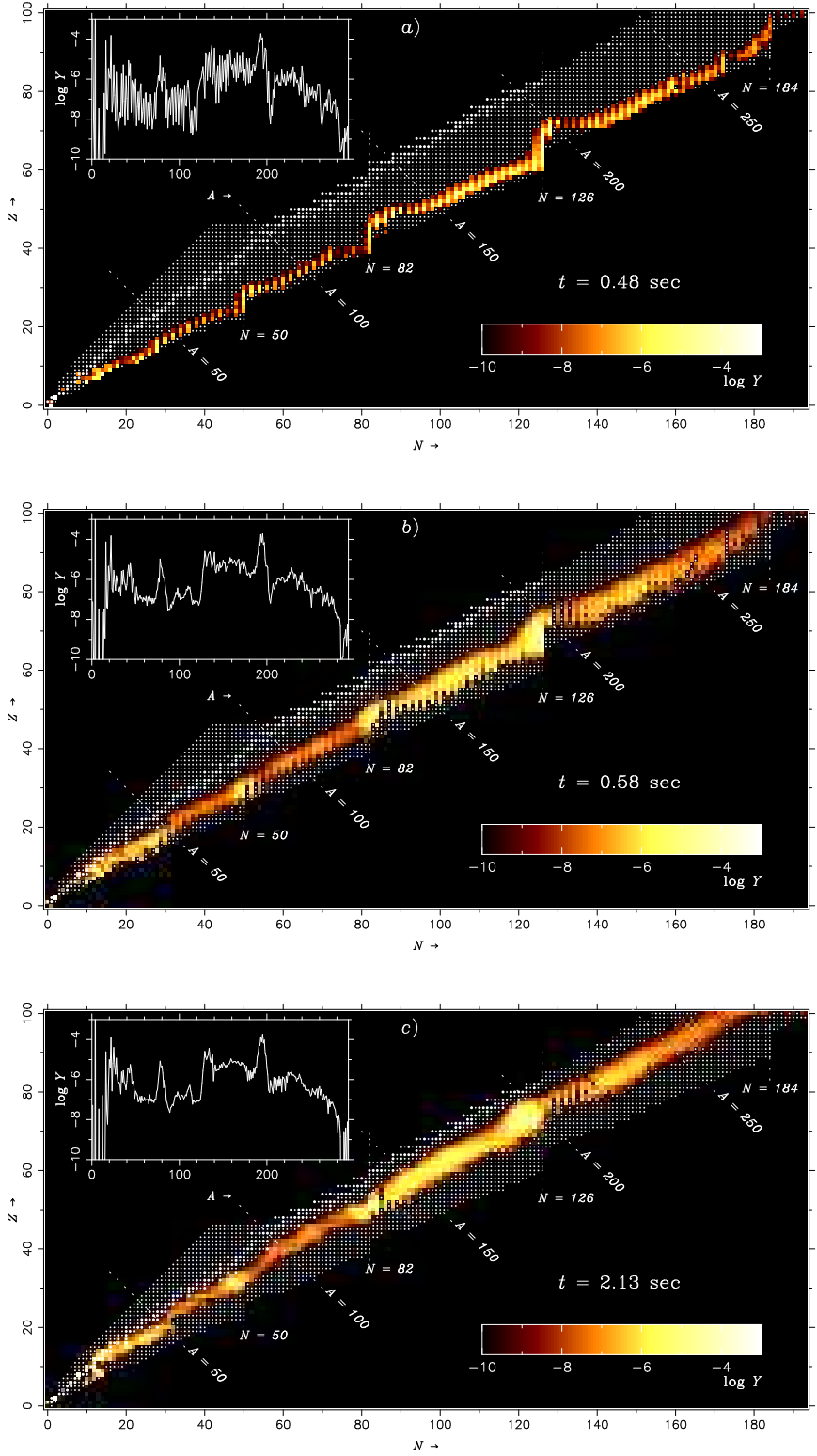}
\caption{Post-freezout evolution of the $r$-process yields at $t =$
(a) 0.48, (b) 0.58, and (c) 2.13 seconds for $\log L_\nu$~(ergs~s$^{-1}) =
51.8$ and $Y_e = 0.43$. The abundances are color-coded in the nuclide chart.
The pattern as a function of mass number is shown in the upper left of the each
panel. The nuclei included in the reaction network are denoted by dots, with
the stable and long-lived isotopes represented by large dots.  }
\end{figure*}

\clearpage
\begin{figure*}
\epsscale{2.0}
\plotone{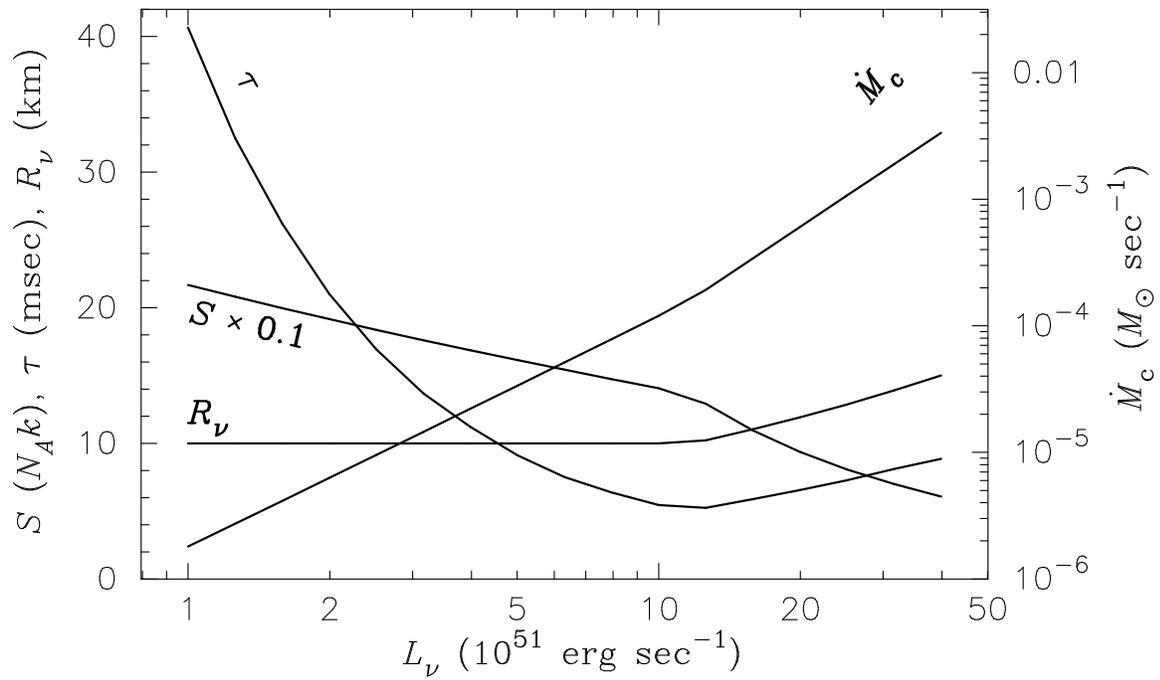}
\caption{The entropy per baryon ($S$), timescale of the seed production
($\tau$), neutrino sphere radius ($R_\nu$), and mass ejection rate ($\dot M_c$)
as functions of the neutrino luminosity ($L_\nu$), in units denoted along the
axes.  }
\end{figure*}

\clearpage
\begin{figure*}
\epsscale{2.0}
\plotone{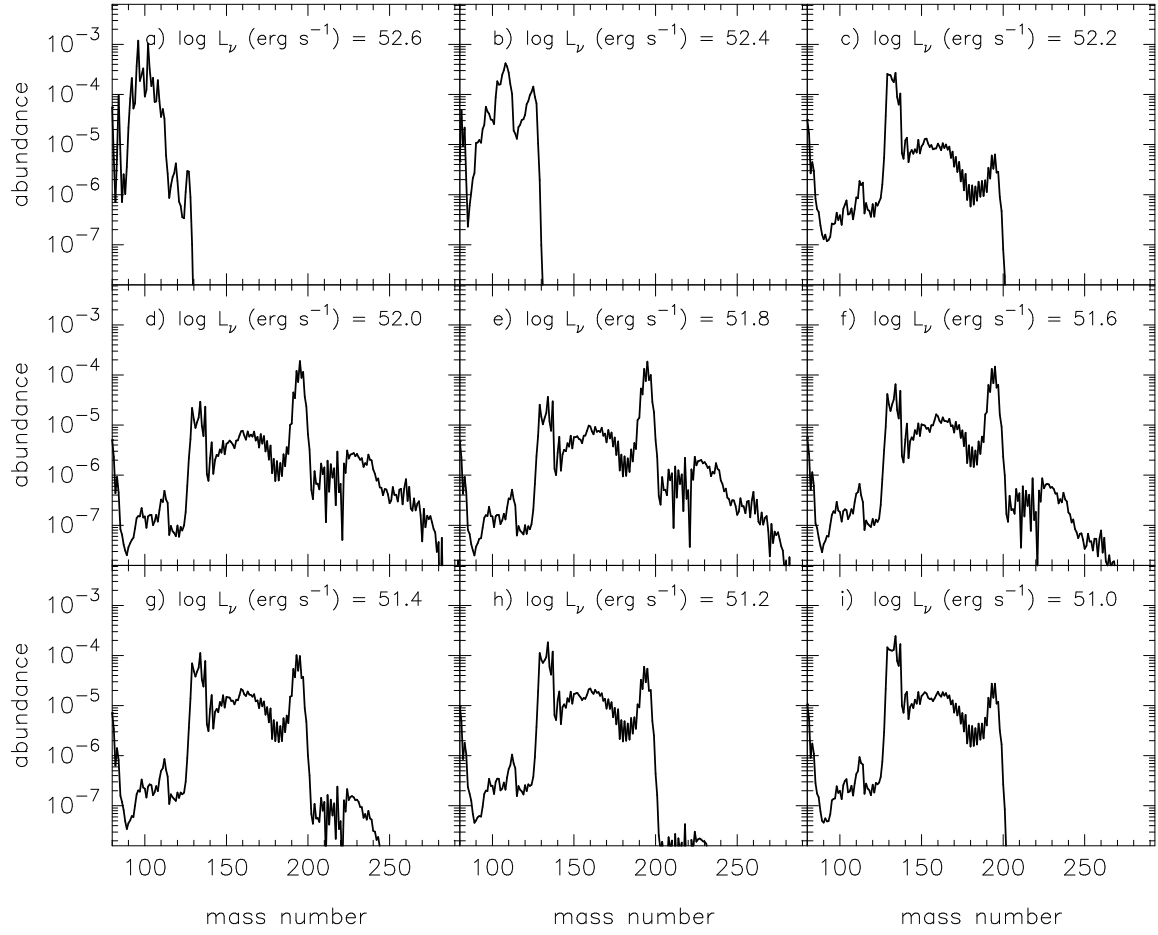}
\caption{Calculated yields for $\log L_\nu$~(ergs~s$^{-1}) = $ (a) 52.6,
(b) 52.4, (c) 52.2, (d) 52.0, (e) 51.8, (f) 51.6, (g) 51.4, (h) 51.2, and
(i) 51.0, with $Y_e = 0.43$}
\end{figure*}

\clearpage
\begin{figure*}
\epsscale{2.0}
\plotone{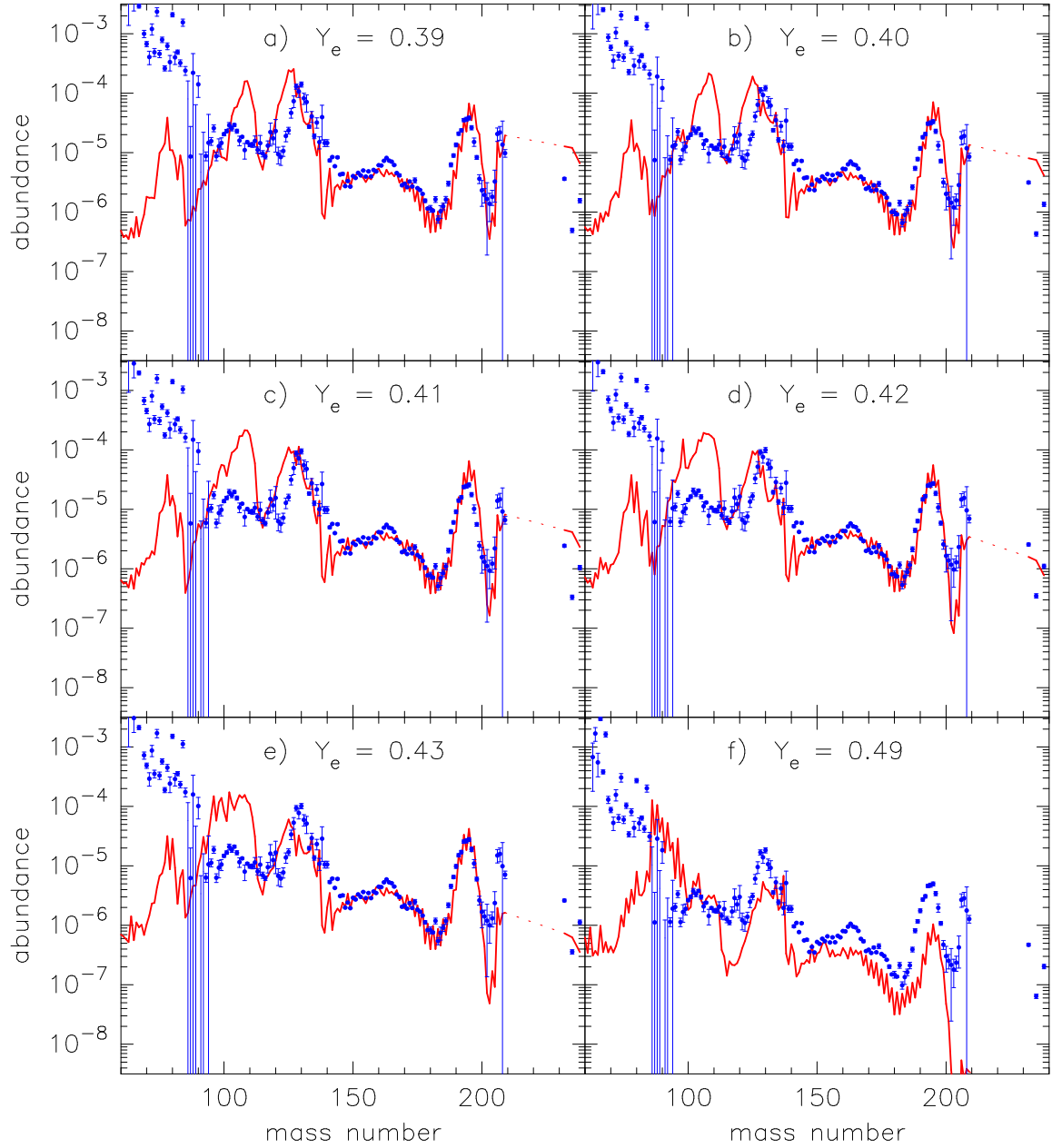}
\caption{Comparison of the mass-integrated yields (line) for $Y_e = $
(a) 0.39, (b) 0.40, (c) 0.41, (d) 0.42, (e) 0.43, and (f) 0.49 with the
solar $r$-pattern (dots) scaled at $^{153}$Eu, as functions of mass number.}
\end{figure*}

\clearpage
\begin{figure*}
\epsscale{2.0}
\plotone{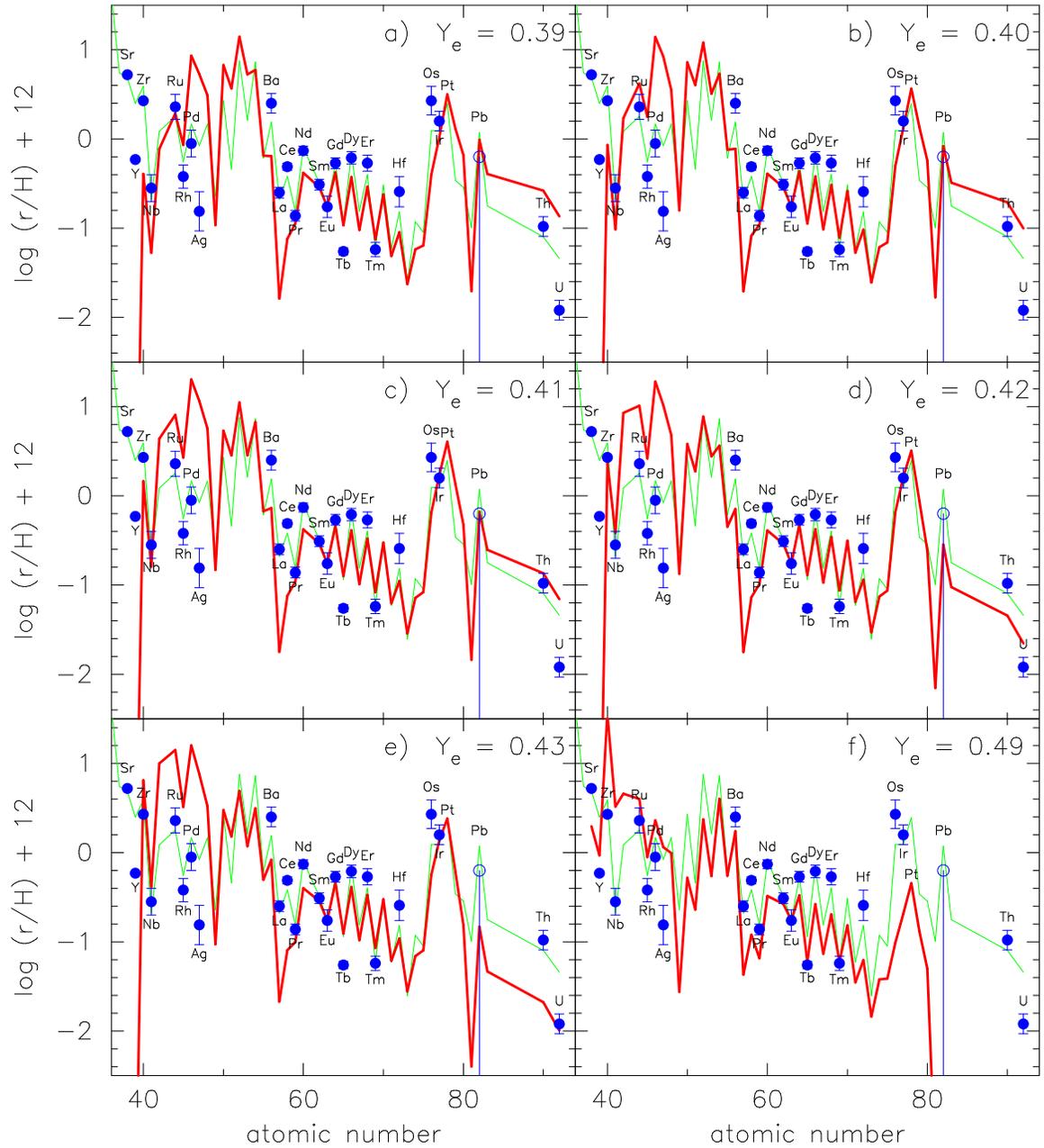}
\caption{Comparison of the mass-integrated yields (the thick line) for $Y_e = $
(a) 0.39, (b) 0.40, (c) 0.41, (d) 0.42, (e) 0.43, and (f) 0.49, scaled at Eu
($Z = 63$), with the abundance pattern of CS~31082-001 (filled circles, with
observational error bars), as functions of atomic number. For Pb, the observed
upper limit is shown by the open circle. The scaled solar $r$-pattern is shown
by the thin line.}
\end{figure*}

\clearpage

\begin{figure*}[t]
\epsscale{2.0}
\plotone{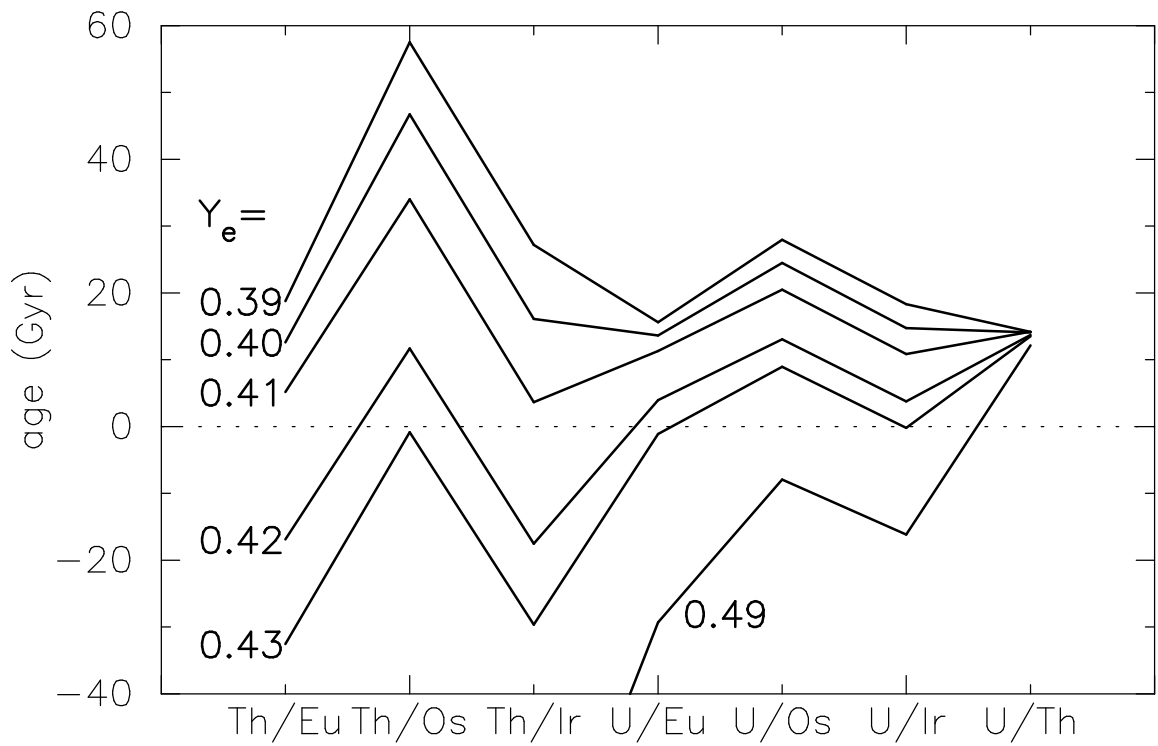}
\caption{Ages of CS~31082-001 derived from various chronometer
 pairs. The robustness of the U-Th pair is clearly shown. The
 superiority of the U-$r$ pairs compared to those of Th-$r$ can also
 be seen.}
\end{figure*}

\end{document}